\documentclass[prl,twocolumn, amsmath,amssymb,
reprint,%
author-year,%
author-numerical,%
dvipdfmx,
superscriptaddress
]{revtex4-1}

\usepackage{graphicx}
\usepackage{bm}
\usepackage{color}

\begin{document}

\title{Breakdown of continuum elasticity due to electronic effects in gold nanotubes}

\author{Shota Ono}
\email{shotaono@muroran-it.ac.jp}
\affiliation{Department of Sciences and Informatics, Muroran Institute of Technology, Muroran 050-8585, Japan}

\author{Hideo Yoshioka}
\affiliation{Department of Physics, Nara Women's University, Nara 630-8506, Japan}

\begin{abstract}
A recent experiment reports a creation of goldene, which is two-dimensional gold with hexagonal structure. By rolling up the goldene, gold nanotubes (GNT) should exist, but their structural and electronic properties are not understood well. Based on first-principles calculations, we demonstrate a breakdown of inverse square law, wherein the curvature energy stored in a GNT decreases with the inverse square of the GNT radius. This is due to the enhanced curvature energy in specific GNTs having nearly flat bands around the Fermi level. We show that the electron states on the flat band of GNT reflect those on the Fermi surface of goldene by using the Bloch and geometric boundary conditions, and that in-plane character of the latter states enhances the curvature energy. 
\end{abstract}

\maketitle

{\it Introduction.---}A physical theory constructed at a macroscopic scale usually breaks down when it is applied to a phenomena at a microscopic scale, where atomic, quantum and/or electronic effects are dominant. The continuum elasticity theory describes the acoustic phonons of low-dimensional systems such as carbon nanotubes (CNT) \cite{suzuura,chico}, twisted bilayer graphene \cite{koshino}, and fullerene-related allotropes \cite{ono2011}, but fails to describe the zone-boundary phonons and optical phonons of those systems \cite{zimm,liu2022,ono2018}. The continuum elasticity theory also describes the periodic rippling of graphene in micrometer scale \cite{bao2009}, but fails to describe it in nanometer scale, where the atomistic treatment considering the $\sigma$ and $\pi$ C-C bonds is necessary \cite{tapaszto}. 

CNT is regarded as a rolled graphene, where the cross-section depends on the chiral indexes $n$ and $m$, and therefore the curvature energy $\Delta E$---a strain energy stored to the nanotube---has been investigated within the continuum elasticity \cite{gulseren,popov}. This is defined as
\begin{eqnarray}
 \Delta E = E(n,m) - E_{\rm 2D},
 \label{eq:DeltaE}
 \end{eqnarray}
where $E(n,m)$ and $E_{\rm 2D}$ are the total energy per atom of the $(n,m)$nanotube and the two-dimensional (2D) monolayer, respectively. The continuum elasticity theory predicts that $\Delta E$ is inversely proportional to the square of radius $R$ \cite{curvature1,curvature2}. 
G\"{u}lseren {\it et al}. have demonstrated that the relation $\Delta E \propto R^{-2}$ holds for zigzag-type CNTs by performing density-functional theory (DFT) calculations \cite{gulseren}. Popov has calculated $\Delta E$ for 187 CNTs with $R=2$ to 15 \AA \ within non-orthogonal tight-binding model and demonstrated that a deviation from the inverse square law is negligibly small \cite{popov}. Surprisingly, the continuum elasticity theory seems to describe the curvature energy of CNTs well.  

Recently, Kashiwaya {\it et al}. have synthesized 2D gold in the hexagonal structure, termed goldene \cite{goldene}, whose stability is predicted by first-principles calculations \cite{yang,koskinen,ono2020}. The goldene is realized by etching away Ti$_3$C$_2$ slabs from Ti$_3$AuC$_2$. The flatness of goldene is due to a hybridization between $6s$ and $5d_{z^2}$ states \cite{Au_cluster}. Recent theoretical calculations have predicted that goldene has low thermal conductivity \cite{goldene_md} and high electrical conductivity \cite{goldene_sigma}, indicating that goldene is useful for miniaturizing electronic devices to the nanoscale. It is interesting to study the energetics of gold nanotubes (GNT) by rolling up goldene. 

GNTs have been studied since the experimental synthesis of helical nanotubes with the chiral index of $(5,3)$ \cite{oshima2003}. 
Several first-principles calculations have been performed for other GNTs \cite{chiral2004,GNT2020,GNT2024}. 
Recently, the electronic structure of achiral GNTs has been calculated for small integers of $n$ and $m$ \cite{GNT2020,GNT2024}.
Due to the metallic property of goldene, the GNTs are also metallic, irrespective to the chiral index. The electron density-of-states (DOS) exhibits a strong peak near the Fermi level, while the appearance of the peak depends on the chirality. In this sense, GNT is different from CNT.

\begin{figure}[b]
\center\includegraphics[scale=0.3]{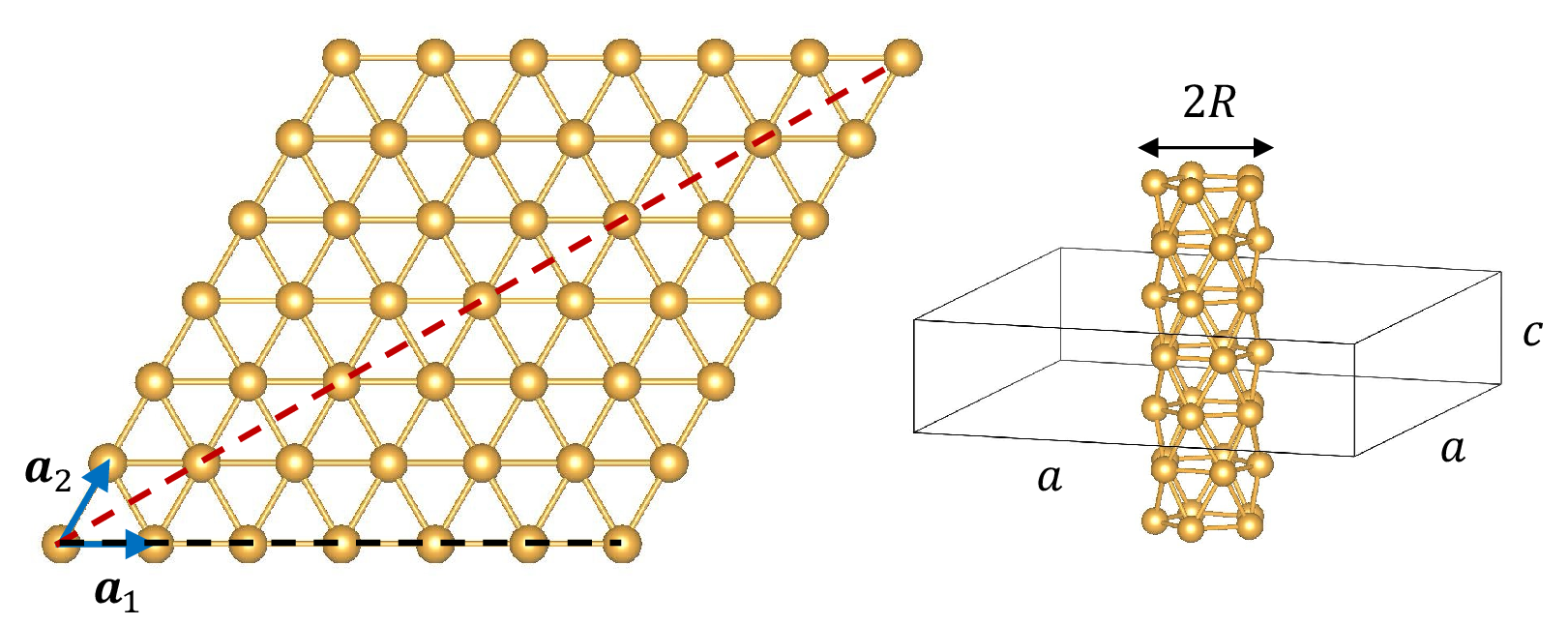}
\caption{Crystal structure of goldene and GNT. Black and red dashed lines are parallel to the chiral vector with indexes $(n,0)$ and $(n,n)$, respectively. } \label{fig1} 
\end{figure}

In this Letter, we demonstrate a breakdown of the inverse square law (ISL) between the curvature energy $\Delta E$ in Eq.~(\ref{eq:DeltaE}) and the GNT radius $R$ by using first-principles approach. A deviation from ISL is significant for ($n$,0)GNTs with $n=5k$ and $5k+2$ ($k$ is a positive integer). This is explained as follows: (i) such ($n$,0)GNTs have nearly flat bands (doubly degenerate) around the Fermi level; (ii) the flat band appears when wave vectors allowed by the Bloch and geometric boundary conditions lie on the Fermi surface with a rounded hexagon shape; and (iii) the electronic states on the Fermi surface possess in-plane character, preferring a planar geometry. In rolling up goldene into such GNTs, bending the in-plane bond needs a large $\Delta E$, giving rise to a deviation from the ISL. The present work suggests that the electronic structure effect is important when continuum elasticity theory is applied to metallic nanostructures.

\begin{figure}
\center\includegraphics[scale=0.55]{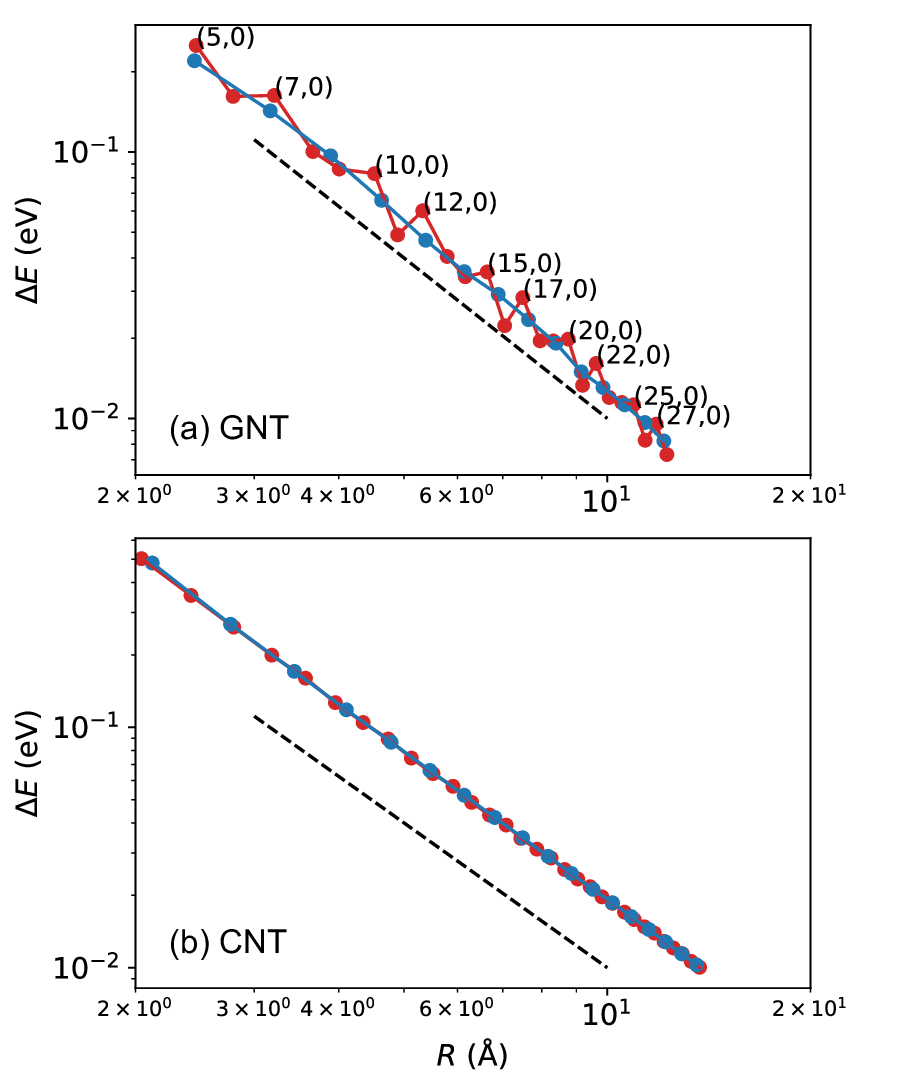}
\caption{The curvature energy $\Delta E$ as a function of $R$ for (a) $(n,0)$ and $(n,n)$ GNTs and (b) $(n,0)$ and $(n,n)$ CNTs. $\Delta E$ of $(n,0)$GNTs shows a significant fluctuation with regard to the ISL (dashed line). The calculated data are plotted for ($n,0$)GNTs with $n=5$-28 (red), ($n,n$)GNTs with $n=3$-16 (blue), ($n,0$)CNTs with $n=5$-35 (red), and ($n,n$)CNTs with $n=3$-20 (blue). } \label{fig2} 
\end{figure}

\begin{figure*}
\center\includegraphics[scale=0.55]{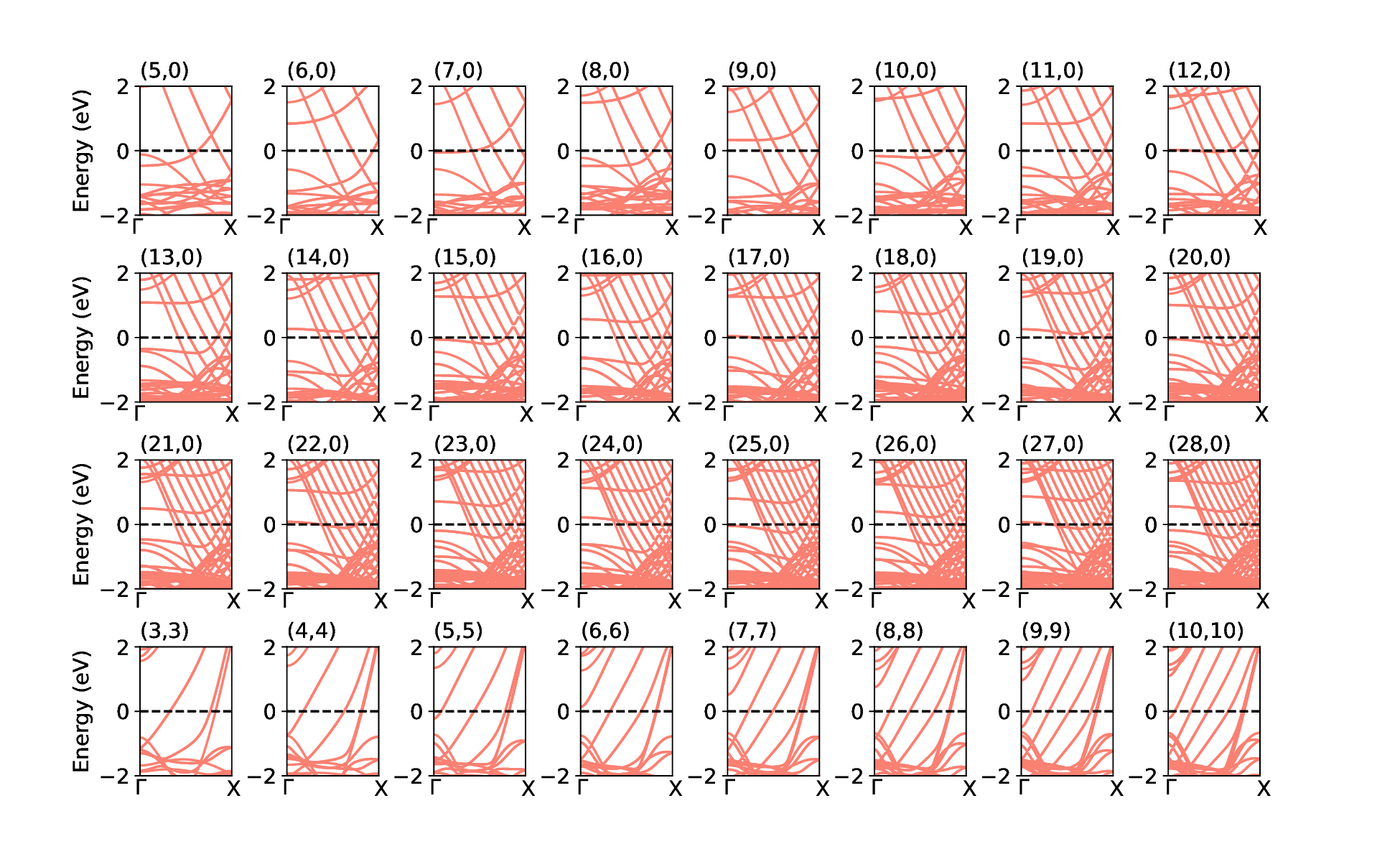}
\caption{Electronic structure of $(n,0)$GNTs from $n=5$ to 28 and $(n,n)$ GNTs from $n=3$ to 10. The energy is measured from the Fermi level. } \label{fig3} 
\end{figure*}

\begin{figure*}
\center\includegraphics[scale=0.5]{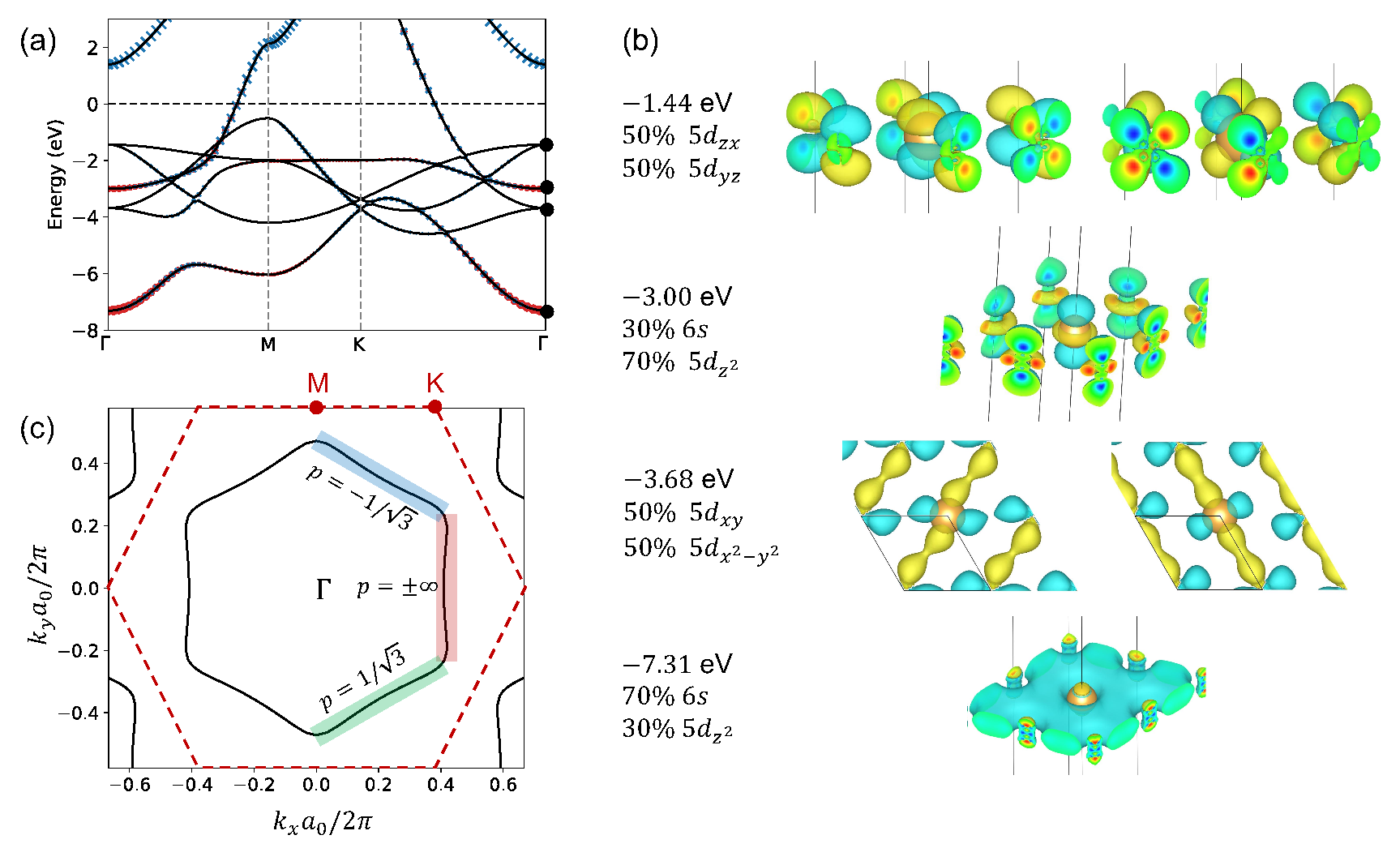}
\caption{(a) Electronic structure of goldene. The contribution from $6s$ and $6p$ states is highlighted by circles (red) and crosses (blue), respectively. The energy is measured from the Fermi level. (b) The Kohn-Sham wavefunction at $\Gamma$ below the Fermi level (black circle in (a)). The percentage for the projection of the Kohn-Sham wavefunctions onto atomic orbitals is also indicated. (c) The Fermi surface of goldene. Dashed line (red) indicates the boundary of the first Brillouin zone. The electronic character is calculated along $k_x = 2\pi/a\times 0.4$ in red shaded area (see main text). 
} \label{fig4} 
\end{figure*}

{\it Continuum elasticity theory.---}We briefly summarize the derivation of $\Delta E \propto R^{-2}$ by following Refs.~\cite{curvature1,curvature2}. We consider a thin film with a thickness $h$ and an area $2\pi RL$, and roll into a hollow cylinder with a length $L$ along the axis and a radius $R$. The total strain energy is expressed by
\begin{eqnarray}
 w = \int_{0}^{2\pi} \frac{M^2}{2YI_z} Rd\theta = \frac{\pi YLh^3}{12R}, 
\end{eqnarray}
where $M = YI_z/R$ is the bending moment, $Y$ is the Young's modulus, and $I_z=Lh^3/12$ is the second moment of area along the axis direction. The total number of atoms is $N=2\pi RL/S_{\rm a}$ with the area per atom $S_{\rm a}$. We thus obtain the curvature energy per atom 
\begin{eqnarray}
 \Delta E = \frac{w}{N} = \frac{Yh^3 S_{\rm a}}{24R^2} = \frac{\alpha}{R^2},
 \label{eq:curvatureE}
\end{eqnarray}
which is inversely proportional to $R^2$. By assuming $Y=80$ GPa for fcc bulk, $h=4.08/\sqrt{3}$ \AA \ for the interlayer distance of [111] direction, and $S_{\rm a}=\sqrt{3}a_{0}^2/2$ with an interatomic distance of $a_0= 2.75$ \AA \ for 2D goldene, $\alpha$ is estimated to be 1.78 eV \AA$^2$. 

{\it Computational details.---}We performed DFT calculations by using Quantum ESPRESSO (QE) \cite{qe}. We used exchange-correlation energy functional within generalized-gradient approximation (GGA) parametrized by Perdew, Burke, and Ernzerhof (PBE) \cite{pbe}. The electron-ion interactions were treated by using ultrasoft pseudopotentials in pslibrary1.0.0 \cite{dalcorso}. The cutoff energies for wavefunction and charge density were set to be 60 Ry and 480 Ry, respectively. The smearing parameter was set to be 0.02 Ry \cite{smearingMV}. The $k$-point distance $\Delta k$ along the $z$-direction is smaller than 0.1 \AA$^{-1}$. To avoid the spurious interaction between the periodic images, we assumed a tetragonal unit cell with the lattice constants of $a=2R+15$ \AA \ and the tube length $c$ (see FIG.~\ref{fig1}). The crystal structure is visualized by using VESTA \cite{vesta}.

The GNT is created by rolling up the goldene. As for the CNT, we define the chiral vector as
\begin{eqnarray}
 \bm{L} = n\bm{a}_1 + m\bm{a}_2,
\end{eqnarray}
where $\bm{a}_1=a_0(1,0)$ and $\bm{a}_2=a_0(1/2,\sqrt{3}/2)$ are the primitive vectors of goldene (see FIG.~\ref{fig1}). $n$ and $m$ are non-negative integers and $n\ge m$. Then, the reciprocal lattice vectors are $\bm{b}_1=2\pi/a_0(1,-1/\sqrt{3})$ and $\bm{b}_2=2\pi/a_0(0,2/\sqrt{3})$. We cut the GNT along the lines perpendicular to the vector $\bm{L}$ through the end points of $\bm{L}$. We connect one edge with the other edge smoothly, and obtain $(n,m)$GNT. The radius is given by $R=a_0\sqrt{n^2+m^2+nm}/(2\pi)$. The cross-section of GNT depends on $(n,m)$ and is classed into straight-type ($n>0$ and $m=0$), zigzag-type ($n=m>0$), and chiral-type (otherwise). The straight- and zigzag-type GNTs have $2n$ atoms in the unit cell. Their positions are expressed as follows: For straight-type GNT,  
\begin{eqnarray}
 & & \left( R\cos (kp), R\sin(kp),0 \right),
 \nonumber\\
 & & \left( R\cos(lp), R\sin(lp),\sqrt{3}a_0/2 \right)
\end{eqnarray}
and for zigzag-type GNT,
\begin{eqnarray}
 & & \left( R\cos (2kp), R\sin(2kp),0 \right),
 \nonumber\\
 & & \left( R\cos (2lp), R\sin(2lp),a_0/2 \right)
\end{eqnarray}
with $k=2j$, $l=2j+1$, $p=\pi/n$, and $j=0,1,\cdots, n-1$. On the other hand, we used atomic simulation environment (ASE) \cite{ase} to prepare the initial geometry of CNT.

Spin polarized calculations were performed in the geometry optimization. We have confirmed that the relaxed structures have no magnetization. 

{\it Curvature energy and electronic structure.---}Figure \ref{fig2} shows $R$-dependence of $\Delta E$ for (a) GNTs and (b) CNTs. 
It is clear that $\Delta E$-$R$ curve of $(n,0)$GNTs (red) is non-monotonic, and $\Delta E$ is large when $n=5k$ and $5k+2$ ($k=1,2,\cdots$). In contrast, $\Delta E$ of $(n,n)$GNTs (blue) decreases with the square of $R$ ($\alpha \simeq 1.3$ eV \AA$^2$ in Eq.~(\ref{eq:curvatureE})). For CNTs, $\Delta E$ also follows the ISL ($\alpha \simeq 2.0$ eV \AA$^2$), irrespective to the chirality (see in FIG.~\ref{fig2}(b)). In this way, the continuum elasticity theory predicting $\Delta E\propto R^{-2}$ fails to describe the energetic stability of $(n,0)$GNTs. We have confirmed that this trend also holds even when the effect of spin-orbit coupling \cite{soc} is included to gold.

Note that $\alpha$ within DFT (1.3 eV \AA$^2$) is smaller than a rough estimation for goldene (1.78 eV \AA$^2$). The former value is obtained if $h$ is decreased to $0.9h$ in Eq.~(\ref{eq:curvatureE}), i.e., the thickness is reduced to 2.12 \AA. This indicates that electron density perpendicular to the goldene surface is decreased due to a lack of Au atoms. Conversely, this implies an increase in the in-plane chemical bonds. 

To understand why only $(n,0)$GNT shows a deviation from the ISL, we calculate the electronic structure of $(n,0)$GNTs. As shown in Fig.~\ref{fig3}, the dispersive bands cross the Fermi level and the number of the dispersive bands increases with $n$. Nearly flat bands (doubly degenerate) are observed around the Fermi level when $n=5k$ and $5k+2$. The appearance of flat bands has also been reported for (7,0)GNT \cite{GNT2024}. Interestingly, these are exactly equal to the $n$s that exhibit an enhancement of $\Delta E$ in Fig.~\ref{fig2}(a). It should be emphasized that neither flat bands nor enhancements of $\Delta E$ are observed in $(n,n)$GNTs (see also FIG.~\ref{fig3}). 

{\it Boundary conditions and electronic character.---}We explain the appearance of flat bands based on the Fermi surface geometry of goldene. The wavevector $\bm{k}$ that the GNT can take is restricted by two boundary conditions. From the Bloch theorem, $\psi (\bm{r}+\bm{L})=e^{i\bm{k}\cdot\bm{L}} \psi (\bm{r})$, where $\psi(\bm{r})$ is the electron wavefunction at $\bm{r}$ in goldene. On the other hand, from the periodic boundary condition for the cylindrical geometry, $e^{i\bm{k}\cdot\bm{L}}=1$. Therefore, one obtains $\bm{k}\cdot\bm{L}=2\pi l$ with integer $l$. For $(n,m)$GNT, 
\begin{eqnarray}
 k_xa_0 \left( n+ \frac{m}{2}\right) +  k_ya_0 \frac{\sqrt{3}m}{2} = 2\pi l. 
 \label{eq:line}
\end{eqnarray}
The electronic band structure around the Fermi level of GNTs is determined by how the straight lines of Eq.~(\ref{eq:line}) cross the Fermi surface of goldene.

Figure \ref{fig4}(a) shows the electronic band structure of goldene. Five $d$-bands are located below the Fermi level. At the $\Gamma$ point, the $6s$ character becomes strong at $-3$ eV, which reflects a hybridization between $6s$ and $5d_{z^2}$ states. This is consistent with previously calculated results for planar Au clusters \cite{Au_cluster}. Figure \ref{fig4}(b) shows the Kohn-Sham wavefunction at $\Gamma$ below the Fermi level. The electron states at the lower two energies, $-7.31$ eV and $-3.68$ eV (doubly degenerate), have in-plane character that should contribute to bending rigidity of goldene.

Figure \ref{fig4}(c) shows the Fermi surface of goldene, showing a rounded hexagon shape. Each vertex is close to the M point, and the slope $p \ (=k_y/k_x)$ of the straight line connecting the nearest vertices is $\pm \infty$ or $\pm 1/\sqrt{3}$. From Eq.~(\ref{eq:line}), the condition of $p=\pm \infty$ is satisfied when $m=0$, i.e., the straight-type GNTs. The condition of $p=\pm 1/\sqrt{3}$ is satisfied when $n=0$ or $n+m=0$, but these cases are excluded due to the definition of $\bm{L}$. The edge of the Fermi surface is located at $k_x \simeq \pm 2\pi/a_0\times 0.4$. When $m=0$, Eq.~(\ref{eq:line}) is reduced to $k_x = 2\pi/a_0\times (l/n)$. In this way, only $(5k,0)$ and $(5k+2,0)$GNTs can have nearly flat bands around the Fermi level, e.g., $2/5=0.4$, $3/7\simeq 0.428$, and $5/12\simeq 0.417$. This is consistent with band structure calculations in Fig.~\ref{fig3}. However, the (5,0)GNT done not have flat bands around the Fermi level possibly due to very small radius ($R=2.46$ \AA). 

We investigate the extent to which the in-plane bond character contributes to the Fermi level. The electronic character on the Fermi line, i.e., $k_x = 2\pi/a_0\times 0.4$ (see FIG.~\ref{fig4}(c)) is as follows: 20\% $6s$, 14\% $6p_x$ and $6p_y$, and 26\% $5d_{xy}$ and $5d_{x^2-y^2}$ states, which are almost independent of $k_y$. It should be emphasized that these electron states have in-plane character. Therefore, in rolling up goldene into $(5k,0)$ and $(5k+2,0)$GNTs having strong degeneracy around the Fermi level, large energy is required to bend the in-plane bond. This causes a failure of the ISL for $(n,0)$GNTs. This scenario is never applied to $(n,n)$GNTs because straight lines of Eq.~(\ref{eq:line}) have $p=-\sqrt{3}$ and cross small portions of the Fermi surface.

We also emphasize that a decrease in $\Delta E$ at $n=6, 8, 9, 11, \cdots$ is understood by the relationship between the Fermi surface and the straight lines in Eq.~(\ref{eq:line}). No Fermi surface is found in the region $0.4 \le \vert k_xa_0/(2\pi)\vert \le 0.6$ for any $k_y$ (see FIG.~\ref{fig4}(c)). For example, when $n=11$, the straight lines, $k_xa_0/(2\pi) = 5/11\simeq 0.45$ and $6/11\simeq 0.54$, go through this region and they are far from the Fermi surface. Therefore, they have little contribution to the bending rigidity, resulting in a decrease in $\Delta E$. 

{\it Conclusions.---}We have demonstrated a breakdown of inverse square law in GNT, wherein the strain energy $\Delta E$ in Eq.~(\ref{eq:DeltaE}) is inversely proportional to $R^2$ by using first-principles approach. The boundary conditions imposed to goldene can yield nearly flat bands around the Fermi level for ($n$,0)GNTs with $n=5k$ and $5k+2$. The electron states on the Fermi surface of goldene possess in-plane character, and therefore $\Delta E$ is enhanced when the goldene is rolled up into ($5k$,0) and ($5k+2$,0)GNTs. We thus postulate an expression for the curvature energy of $\Delta E=\alpha/R^2+\beta_{\rm el}$, i.e., a sum of the elastic energy and the energy correction due to electronic structure. The present work highlights an importance of $\beta_{\rm el}$ when considering nanoscale deformations of metallic monolayers. 

\begin{acknowledgments}
This work was supported by JSPS KAKENHI (Grant No. 21K04628 and No. 24K01142). Calculations were done using the facilities of the Supercomputer Center, the Institute for Solid State Physics, the University of Tokyo.
\end{acknowledgments}




\end{document}